**Eigen power-flow-density vortices of magnetostatic modes in thin ferrite disks**

M. Sigalov, E. O. Kamenetskii, and R. Shavit

Ben-Gurion University of the Negev, Beer Sheva 84105, Israel

April 9, 2008

In confined magnetically ordered structures one can observe vortices of magnetization and electromagnetic power flow vortices. There are topologically distinct and robust states. In this paper we show that in a normally magnetized quasi-2D ferrite disk there exist eigen power-flow-density vortices of magnetic-dipolar-mode oscillations. Because of such circular power flows, the oscillating modes are characterized by stable magnetostatic energy states and discrete angular moments of the wave fields. We show that the power-flow-density vortices of magnetostatic modes can be excited by electromagnetic fields of a microwave cavity. There is a clear correspondence between the power-flow-density vortex structures in a ferrite disk derived from an analytical solution of the magnetostatic-wave spectral problem and obtained by the numerical-simulation electromagnetic program.

PACS numbers: 76.50.+g, 68.65.-k, 03.75.Lm

----------------------

In spite of the fact that vortices are observed in different kinds of physical phenomena, yet such "swirling" entities seem to elude an all-inclusive definition. In quite a number of problems one can define a vortex as a circular flow which is attributed with a certain phase factor and a circular integral of a gradient of the phase gives a non-zero quantity. This quantity is multiple to a number of full rotations.

In ferromagnetic systems, one can clearly distinguish three characteristic scales. There are the scales of the spin (exchange interaction) fields, the magnetostatic (dipole-dipole interaction) fields, and the electromagnetic fields. These characteristic scales may define different vortex states. Together with magnetization vortices in micrometer- or submicrometer-size ferromagnetic samples [1, 2] and the magnetostatic (MS) vortex behaviors in thin ferrite films [3], one can observe electromagnetic vortices originated from ferrite samples in microwave cavities [4 – 7]. In the last case, vortices appear because of the time-reversal-symmetry-breaking (TRSB) effects resulting in a rich variety of the electromagnetic wave topological phenomena [4 – 7].

In their studies, Boardman *et al* [3] showed that for MS waves excited by three planar antennas in a normally magnetized ferrite film one can form a stationary linear phase defect structure resulting in appearance of the power-flow-density vortices. There are *induced* MS-wave vortices. In this paper we show that there is a possibility to obtain *eigen* power-flow-density vortices of MS oscillations which appear due to special topological effects in a normally magnetized quasi-2D ferrite disk. Based on numerical simulations, we show that the eigen power-flow-density vortices of MS modes in a ferrite disk can be excited by electromagnetic fields of a microwave cavity. While for thick ferrite disks studied in [5 – 7] no multiresonance absorption spectra were observed in a frequency range of the cavity resonance, in a case of a quasi-2D ferrite disk with MS modes one obtains eigenstates of the vortex structures.

The magnetic dipole interaction provides us with a long-range mechanism of interaction, where a magnetic medium is considered as a continuum. It is well known that MS ferromagnetism has a character essentially different from exchange ferromagnetism [8, 9]. This statement finds a strong confirmation in confinement phenomena of magnetic-dipolar-mode (MDM) (or MS) oscillations. A recently published spectral theory of MDMs in quasi-2D ferrite disks [10 – 12] gives a deep insight into an explanation of the experimental multiresonance line absorption spectra shown both in well known previous [13, 14] and new [15, 16] studies. This theory suggests an existence of othogonality relations for the MS-potential eigen wave functions [10, 11] and dynamical symmetry breaking effects for MDM oscillations [12]. One of the most attractive aspects of the symmetry breaking effects in MDM oscillations is the presence of the vortex states which appears due to special boundary conditions on a lateral surface of a normally magnetized quasi-2D ferrite disk [12]. Based on the MS spectral problem solutions, in this paper we show that the border states of MDM oscillations lead to eigen MS power-flow-density vortices in a ferrite disk. Due to these circular eigen power flows, the MDMs are characterized by stable MS energy states.

For MDMs in a normally magnetized ferrite disk, circular flows of power densities are attributed with the phase factors of MS-potential wave functions. For monochromatic fields with time variation $\sim e^{i\omega t}$ the power flow density for a certain magnetic dipolar mode $n$ is expressed in Gaussian units as [12]:

$$\vec{p}_n = \frac{i\omega}{16\pi}\left(\psi_n^* \vec{B}_n - \psi_n \vec{B}_n^*\right), \tag{1}$$

where $\psi_n$ is the MS-potential wave function, $\vec{B}_n = -\overleftrightarrow{\mu} \cdot \vec{\nabla}\psi_n$, and $\overleftrightarrow{\mu}$ is the permeability tensor. In a normally magnetized (with a normal directed along $z$ axis) quasi-2D ferrite disk, the mode fields can be represented as [10 – 12]: $\psi_n = C_n \xi_n(z)\widetilde{\varphi}_n(x,y)$, $\vec{B}_n = (B_n)_z \vec{e}_z + (B_n)_\perp \vec{e}_\perp$, where $(B_n)_z = -C_n \dfrac{\partial \xi_n(z)}{\partial z}\widetilde{\varphi}_n(x,y)$ and $(B_n)_\perp = -C_n \xi_n(z)\left[\overleftrightarrow{\mu}_\perp \cdot \vec{\nabla}_\perp \widetilde{\varphi}_n(x,y)\right] \cdot \vec{e}_\perp$, $\xi_n(z)$ is an amplitude factor, $C_n$ is a dimensional coefficient, and $\widetilde{\varphi}_n(x,y)$ is a dimensionless membrane function for mode $n$. Subscript $\perp$ corresponds to transversal (with respect to $z$ axis) components. In a cylindrical coordinate system, it is easy to show that for oscillating MDMs in a quasi-2D ferrite disk, the $z$ and $r$ components of the power flow density are equal to zero. There is the only real azimuth component:

$$(p_n)_\theta = \frac{i\omega}{16\pi}C_n^2(\xi(z))^2\left[-\mu\frac{1}{r}\left(\widetilde{\varphi}_n^* \frac{\partial \widetilde{\varphi}_n}{\partial \theta} - \widetilde{\varphi}_n \frac{\partial \widetilde{\varphi}_n^*}{\partial \theta}\right) + i\mu_a\left(\widetilde{\varphi}_n^* \frac{\partial \widetilde{\varphi}_n}{\partial r} + \widetilde{\varphi}_n \frac{\partial \widetilde{\varphi}_n^*}{\partial r}\right)\right]. \tag{2}$$

The total MS-potential membrane function $\widetilde{\varphi}$ is represented as a product of two functions [12]:

$$\widetilde{\varphi} = \widetilde{\eta}(r,\theta)\,\delta_\pm. \tag{3}$$

Function $\widetilde{\eta}(r,\theta)$ is a single-valued membrane function written as

$$\widetilde{\eta}(r,\theta) = R(r)\phi(\theta), \tag{4}$$

where $R(r)$ is described by the Bessel functions and $\phi(\theta) \sim e^{-i\nu\theta}$, $\nu = \pm 1, \pm 2, \pm 3....$ Function $\delta_\pm$ is a double-valued (spin-coordinate-like) function, which is represented as $\delta_\pm \equiv f_\pm e^{-iq_\pm\theta}$, where $q_\pm = \pm\dfrac{1}{2}$. For amplitudes $f$ we have $f_+ = -f_-$ with normalization $\left|f_\pm\right| = 1$.

Circular flows of the power density in a MDM ferrite disk are attributed with phase factors of a MS-potential wave function. The MDM topological effects are manifested through the generation of relative phases which accumulate on the boundary wave functions $\delta_{\pm}$. Due to this function one has a phase factor which defines a power-flow-density vortex of a MDM. With use of Eq. (4) one rewrites Eq. (2) as

$$\left(p_n(r)\right)_\theta = \frac{R_n(r)}{8\pi}\,\omega\,C_n^2\left(\xi_n(z)\right)^2\left[-R_n(r)\frac{\mu}{r}(\nu_n) - \mu_a\frac{\partial R_n(r)}{\partial r}\right]. \qquad (5)$$

There is non-zero circulation quantities $\left(p_n(r)\right)_\theta$ around a circle $2\pi r$. An amplitude of a MS-potential function is equal to zero at $r = 0$. For a scalar wave function, this presumes the Nye and Berry phase singularity [17]. Circulating quantities $\left(p_n(r)\right)_\theta$ are the MDM power-flow-density vortices with cores at the disk center. At a vortex center amplitude of $\left(p_n\right)_\theta$ is equal to zero. It follows from Eq. (5) that for a given mode number $n$ characterizing by a certain Bessel function $R_n(r)$ there will be different functions of the power flow density $\left(p_n(r)\right)_\theta$ for different signs of the azimuth number $\nu_n$.

To find functions $\xi(z)$ and $R(r)$ we have to solve a system of the following two equations [11]:

$$\tan\left(\beta^{(F)}h\right) = -\frac{2\sqrt{-\mu}}{1+\mu} \qquad (6)$$

and

$$(-\mu)^{\frac{1}{2}}\frac{J_\nu'}{J_\nu} + \frac{K_\nu'}{K_\nu} = 0, \qquad (7)$$

corresponding to the so-called essential boundary conditions well known in variational methods [18]. This gives the energy orthogonality relations for magnetic-dipolar modes $\tilde{\eta}(r,\theta)$ [11, 12]. Here $h$ and $\Re$ are, respectively, a thickness and a radius of a ferrite disk, $\beta^{(F)}$ is the wave number of a MS wave propagating in a ferrite along a bias magnetic field, $J_\nu, J_\nu', K_\nu,$ and $K_\nu'$ are the values of the Bessel functions of order $\nu$ and their derivatives (with respect to the argument) on a lateral cylindrical surface ($r = \Re,\ 0 \leq z \leq h$).

In further analysis we consider MDMs having fundamental thickness and first-order-azimuth distributions. Numbers $n$ in Eq. (5) correspond to different radial variations. Fig. 1 gives the calculated distributions of $\left(p_n\right)_\theta$ for first two modes ($n = 1, 2$) at $\nu_n = +1$ when a bias magnetic field is directed along $z$ axis. These distributions clearly show the power-flow-density vortices. For our calculations we used a lossless normally magnetized ferrite disk with diameter $2\Re = 3$ mm and thickness $d = 0.05$ mm. The ferrite saturation magnetization is $4\pi M_0 = 1880$ G and a bias magnetic field is $H_0 = 4900$ Oe. One can see that there are eigen power-flow-density vortices with very different topological structures.

It follows, however, that an analysis of excitation of these power-flow-density vortices by external electromagnetic fields is beyond the frames of any analytical solutions. Because of the TRSB effects, a system of a cavity with an embedded inside ferrite disk (even having sizes much small compared with the cavity sizes) is not a weakly perturbed integrable system, but a non-integrable system [5 – 7]. Based on the HFSS-program numerical studies [19], we analyze excitation of the power-flow-density vortices in a ferrite disk placed in a microwave cavity. For our numerical studies we used a short-wall rectangular waveguide section. The disk axis was oriented along the E-field of a waveguide $TE_{10}$ mode. The disk parameters were the same as for the above analytical calculations. Additionally, we took into account the material losses as the linewidth $\Delta H = 0.8\,\text{Oe}$. This corresponds to the parameters used in experiments [16]. Fig. 2 shows numerically obtained frequency characteristic of an absorption coefficient for a ferrite disk in a waveguide cavity. One clearly sees the multiresonance absorption spectra. The peak amplitudes reflect the fact of different waveguide field structures at different frequencies of the disk modes. In Fig. 2 we also show the resonance peak positions obtained from an analytical solution of Eqs. (6) and (7). There is a very good correlation between the analytical and numerical peak positions. For numerically obtained modes, one can observe topological resonant states. Every resonant state is characterized by a strong pronounced eigenfunction pattern with a topologically distinct vortex structure. As an example, Fig. 3 shows a typical gallery of the magnetic field distributions on the upper plane of a ferrite disk for the second mode at different time phases. A very peculiar property of these pictures is the fact of the azimuthal rotation of the mode magnetic field. When the transverse mode is transformed following a closed path in the space of modes, the phase of the final mode state differs from that of the initial state by $\phi = \phi_d + \phi_g$, where $\phi_d$ and $\phi_g$ are the dynamical and geometrical phases, respectively [20]. In a supposition of possible analytical description, the modes with $4\pi$ azimuthal rotation should be represented by double-valued functions.

Based on numerical studies, we can represent the Poynting-vector distributions inside a ferrite disk corresponding to the observed resonant states. Fig. 4 gives such distributions for first two modes ($n = 1, 2$). There are the power-flow vortices. The black arrows clarify the power-flow directions inside a disk for every given mode. One can find a very good correspondence between the numerical-simulation Poynting-vector vortices in Fig. 4 and analytically calculated eigen MDM power-flow-density vortices shown in Fig. 1. When comparing the vortices in Figs. 1 and 4, one should take into account the fact that in the numerical-simulation analysis, the ferrite material losses were taken into consideration. This presumes certain diffusion of the vortex pictures in Fig. 4.

We showed that in a normally magnetized quasi-2D ferrite disk there exist eigen power-flow-density vortices of MDM oscillations. Based on the HFSS-program numerical studies, we showed possibility of excitation of the MDM power-flow-density vortices in a ferrite disk placed in a microwave cavity. The pictures of the vortex structures obtained from the numerical-simulation electromagnetic program are in a good correspondence with the analytically derived eigen MS power-flow-density vortices in a ferrite disk.

-----------

[1] T. Shinjo, T. Okuno, R. Hassdorf, K. Shigeto, and T. Ono, Science **289**, 930 (2000).
[2] K. Yu. Guslienko, V. Novosad, Y. Otani, H. Shima, and K. Fukamichi, Phys. Rev. B **65**, 024414 (2002).
[3] A. D. Boardman, Yu. G. Rapoport, V. V. Grimalsky, B. A. Ivanov, S. V. Koshevaya, L. Velasco, and C. E. Zaspel, Phys Rev. **71**, 026614 (2005).
[4] M. Vraničar, M. Barth, G. Veble, M. Robnik, and H.-J. Stöckmann, J. Phys. A: Math. Gen. **35**, 4929 (2002).
[5] E. O. Kamenetskii, M. Sigalov, and R. Shavit, Phys. Rev. E **74**, 036620 (2006).
[6] M. Sigalov, E. O. Kamenetskii, and R. Shavit, Phys. Lett. A **372**, 91 (2008).
[7] M. Sigalov, E. O. Kamenetskii, and R. Shavit, J. Appl. Phys. **103**, 013904 (2008).
[8] J. M. Luttinger and L. Tisza, Phys. Rev. **70**, 954 (1946).
[9] H. Puszkarski, M. Krawczyk, and J.-C. S. Levy, Phys. Rev. B **71**, 014421 (2005).
[10] E. O. Kamenetskii, Phys. Rev. E **63**, 066612 (2001).
[11] E. O. Kamenetskii, M. Sigalov, and R. Shavit, J. Phys.: Condens. Matter **17**, 2211 (2005).

[12] E. O. Kamenetskii, J. Phys. A: Math. Theor. **40**, 6539 (2007).

[13] J. F. Dillon Jr., J. Appl. Phys. **31**, 1605 (1960).

[14] T. Yukawa and K. Abe, J. Appl. Phys. **45**, 3146 (1974).

[15] E. O. Kamenetskii, A.K. Saha, and I. Awai, Phys. Lett. A **332**, 303 (2004).

[16] M. Sigalov, E. O. Kamenetskii, and R. Shavit, arXiv:0712.2305.

[17] J. F. Nye and M. V. Berry, Proc. R. Soc. London Ser. A **336**, 165 (1974).

[18] S. G. Mikhlin, *Variational Methods in Mathematical Physics* (New York, McMillan, 1964).

[19] High Frequency Structure Simulator, www.hfss.com, Ansoft Corporation, Pittsburgh, PA 15219.

[20] E. J. Galvez, P. R. Crawford, H. I. Sztul, M. J. Pysher, P. J. Haglin, and R. E. Williams, Phys Rev. Lett. **90**, 203901 (2003).

**Figure captions**

Fig. 1. Power-flow-density vortices for magnetic-dipolar modes in a ferrite disk for first two modes. Disk diameter $2\Re = 3\,\text{mm}$.

Fig. 2. Spectrum of MDMs in a ferrite disk. (a) A numerically obtained frequency characteristic of an absoption coefficient for a ferrite disk in a waveguide cavity; an insertion shows a disk position in a cavity. (b) Analytically calculated MDM resonance peak positions.

Fig. 3. A gallery of the magnetic field distributions on the upper plane of a ferrite disk for the second mode at different time phases. There is the mode with $4\pi$ azimuthal rotation.

Fig. 4. The Poynting-vector distributions inside a ferrite disk corresponding to the topological resonant states. The black arrows clarify the power-flow directions inside a disk for every given mode.

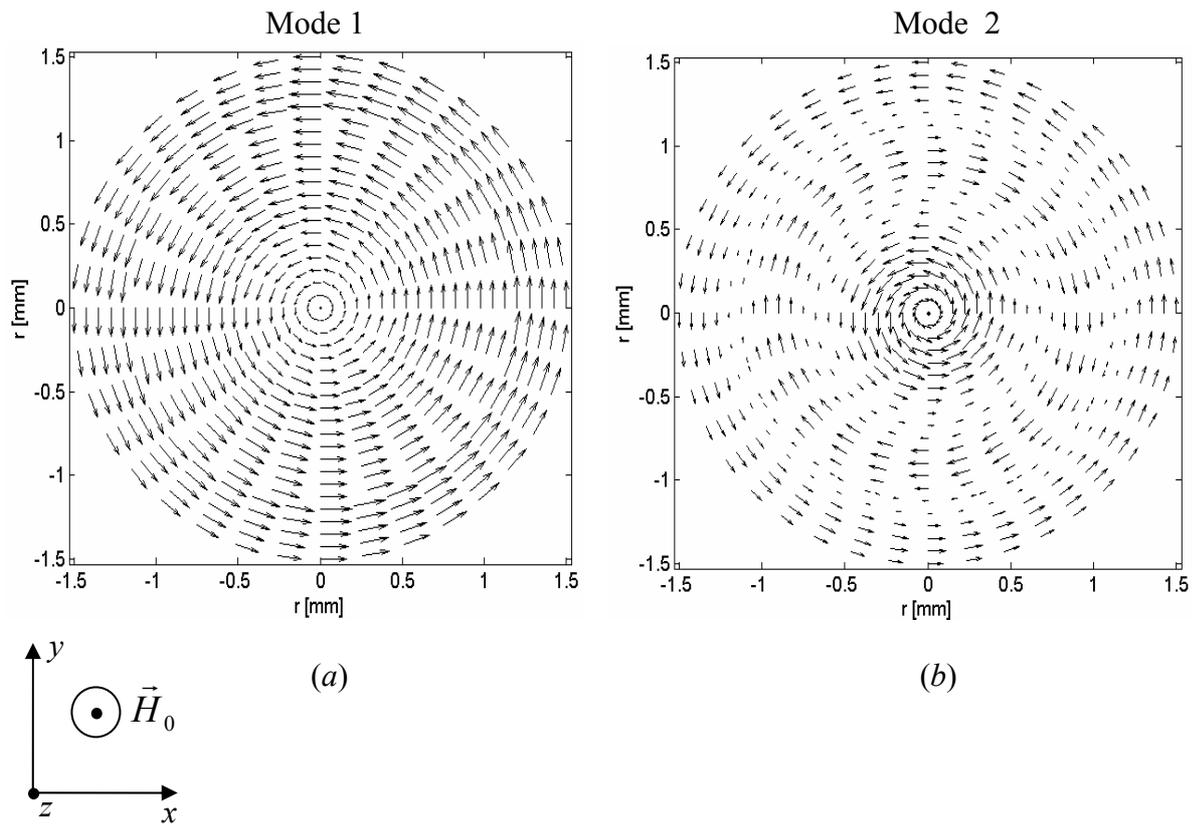

Fig. 1. Power-flow-density vortices for magnetic dipolar modes in a ferrite disk for first two modes. Disk diameter $2\Re = 3\,\text{mm}$.

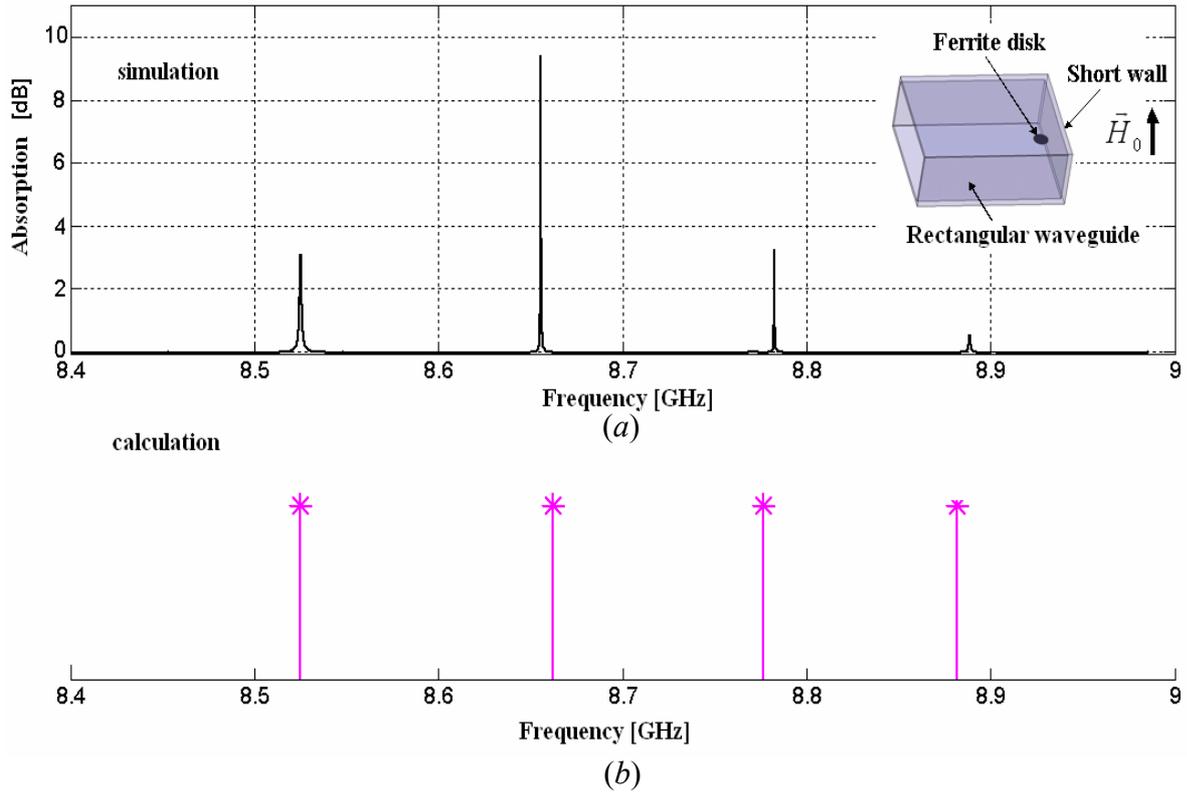

Fig. 2. Spectrum of MDMs in a ferrite disk. (a) A numerically obtained frequency characteristic of an absorption coefficient for a ferrite disk in a waveguide cavity; an insertion shows a disk position in a cavity. (b) Analytically calculated MDM resonance peak positions.

Mode 2

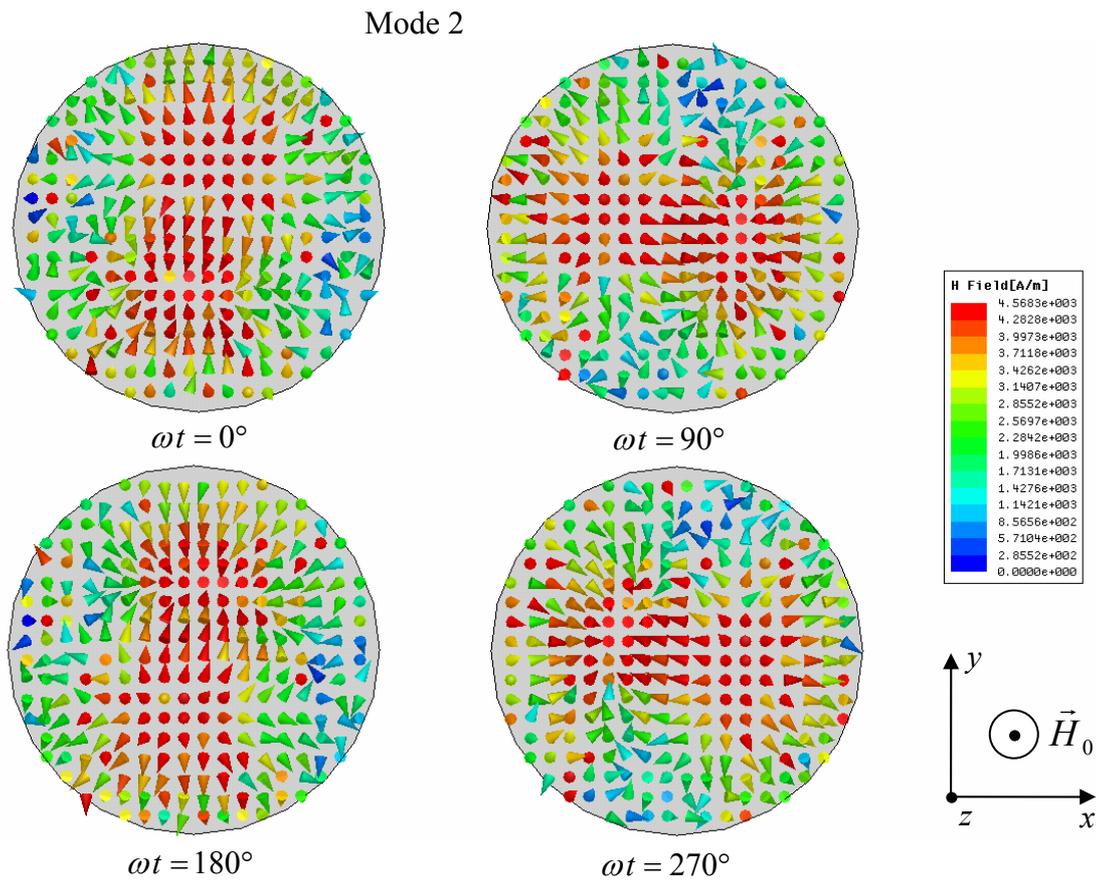

$\omega t = 0°$

$\omega t = 90°$

$\omega t = 180°$

$\omega t = 270°$

Fig. 3. A gallery of the magnetic field distributions on the upper plane of a ferrite disk for the second mode at different time phases. There is the mode with $4\pi$ azimuthal rotation.

Mode 1

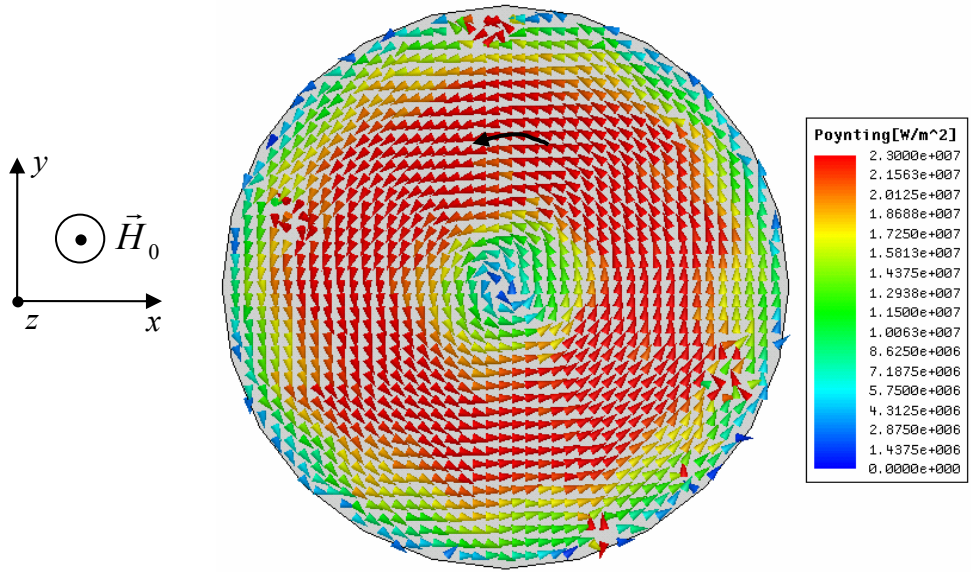

(*a*)

Mode 2

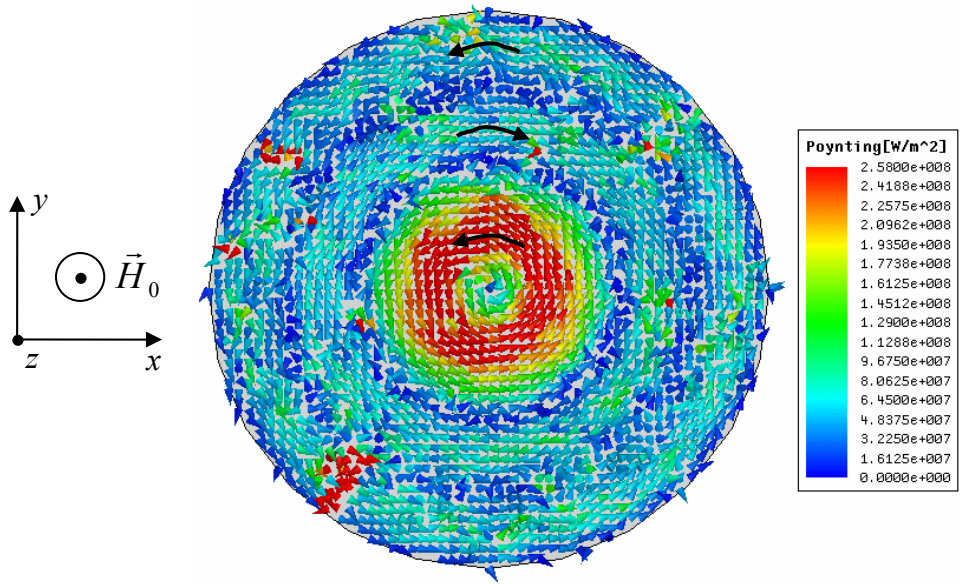

(*b*)

Fig. 4. The Poynting-vector distributions inside a ferrite disk corresponding to the topological resonant states. The black arrows clarify the power-flow directions inside a disk for every given mode.